OPEN

# Fluorine doping: a feasible solution to enhancing the conductivity of high-resistance wide bandgap $Mg_{0.51}Zn_{0.49}O$ active components

Lishu Liu[1], Zengxia Mei[1], Yaonan Hou[1], Huili Liang[1], Alexander Azarov[2], Vishnukanthan Venkatachalapathy[2], Andrej Kuznetsov[2] & Xiaolong Du[1]

**N-type doping of high-resistance wide bandgap semiconductors, wurtzite high-Mg-content $Mg_xZn_{1-x}O$ for instance, has always been a fundamental application-motivated research issue. Herein, we report a solution to enhancing the conductivity of high-resistance $Mg_{0.51}Zn_{0.49}O$ active components, which has been reliably achieved by fluorine doping via radio-frequency plasma assisted molecular beam epitaxial growth. Fluorine dopants were demonstrated to be effective donors in $Mg_{0.51}Zn_{0.49}O$ single crystal film having a solar-blind 4.43 eV bandgap, with an average concentration of $1.0 \times 10^{19}$ F/cm$^3$.The dramatically increased carrier concentration ($2.85 \times 10^{17}$ cm$^{-3}$ vs ~$10^{14}$ cm$^{-3}$) and decreased resistivity (129 $\Omega \cdot$ cm vs ~$10^6$ $\Omega$ cm) indicate that the electrical properties of semi-insulating $Mg_{0.51}Zn_{0.49}O$ film can be delicately regulated by F doping. Interestingly, two donor levels (17 meV and 74 meV) associated with F were revealed by temperature-dependent Hall measurements. A Schottky type metal-semiconductor-metal ultraviolet photodetector manifests a remarkably enhanced photocurrent, two orders of magnitude higher than that of the undoped counterpart. The responsivity is greatly enhanced from 0.34 mA/W to 52 mA/W under 10 V bias. The detectivity increases from $1.89 \times 10^9$ cm Hz$^{1/2}$/W to $3.58 \times 10^{10}$ cm Hz$^{1/2}$/W under 10 V bias at room temperature. These results exhibit F doping serves as a promising pathway for improving the performance of high-Mg-content $Mg_xZn_{1-x}O$-based devices.**

Wide bandgap oxide semiconductors such as $\beta$-$Ga_2O_3$, $CeO_2$, ZnO and its alloy, $Mg_xZn_{1-x}O$, have been more and more attractive for applications in ultraviolet (UV) photodetectors (PDs), photocatalyst, transparent conductive electrodes, and electronic devices etc[1–4]. Among them, wurtzite $Mg_xZn_{1-x}O$ (W-$Mg_xZn_{1-x}O$) is highlighted with a combination of large, theoretically tunable bandgap (3.37–6.3 eV), low growth temperature (100–450 °C), capabilities of wet-etch processing, etc. The environment-friendly and biocompatible characteristics also make MgZnO appealing for UV device applications. Moreover, ZnO is remarkably resistant to high-energy particle irradiation, which is extremely important for UV PDs working in the outer space[5]. However, one of the biggest challenges is how to reproducibly synthesize high-quality single-phase W-$Mg_xZn_{1-x}O$ with high Mg content. The well-known phase segregation problem in MgZnO[6] makes it difficult to extend the cutoff wavelength into the significant solar-blind UV spectral region. In our previous work, solar-blind 4.55 eV bandgap W-$Mg_{0.55}Zn_{0.45}O$ components have been fabricated on c-sapphire by applying a unique interface engineering technique[2,7]. Ju *et al.*[8] focused on synthesizing cubic MgZnO (C-MgZnO) and successfully fabricated the solar-blind UV detector, which opens up an important research direction on deep UV materials.

[1]Key Laboratory for Renewable Energy, National Laboratory for Condensed Matter Physics, Institute of Physics, Chinese Academy of Sciences, P.O. Box 603, Beijing 100190, China. [2]Department of Physics, University of Oslo, P.O. Box 1048 Blindern, NO-0316 Oslo, Norway. Correspondence and requests for materials should be addressed to Z.M. (email: zxmei@iphy.ac.cn) or X.D. (email: xldu@iphy.ac.cn)

Another crucial issue restricting the practical use of high-Mg-content W-$Mg_xZn_{1-x}O$ is its notably high resistance[9–11]. For example, the single-crystal W-$Mg_{0.55}Zn_{0.45}O$ film[7] exhibits such high resistivity that the solar-blind UV PDs[2,4] fabricated on this film demonstrate a ~5 nA photocurrent when biased at 150 V and under 254 nm UV light illumination[2], which is insufficient for practical requirements. Tuning the conductivity is therefore specifically necessary for high-Mg-content W-$Mg_xZn_{1-x}O$ films and related devices. By intentionally introducing point defects—Zn interstitial, Liu *et al.*[12] reported their interesting result on the largely decreased resistivity of MgZnO as low as 0.053 Ω · cm. Meanwhile, heterovalent cation dopants–$Ga^{3+}$ and $Al^{3+}$ for instance–have been added into W-$Mg_xZn_{1-x}O$ to create electron carriers. However, the effectiveness of these dopants as donors appears to decrease drastically as the Mg content (x) in $Mg_xZn_{1-x}O$ increases[13,14], similar to the case of Si in $Al_xGa_{1-x}N$[15]. Even more worse, such dopants like Ga might cause phase segregation in low-Mg-content W-$Mg_xZn_{1-x}O$ ($x \leq 0.2$) films[16]. Recently, Guo *et al.*[17] reported a quaternary alloy of $Zn_{0.9}Mg_{0.1}OF_{0.03}$ to be potentially applied as transparent electrodes. There has been no report on tuning the electrical properties of high-Mg-content W-MgZnO with bandgap in solar-blind range yet. Therefore, it is worth exploring some new methods for effective n-type doping in W-$Mg_xZn_{1-x}O$ ($x > 0.4$) films in order to promote the corresponding device performance.

In this work, via comparative studies of doping with different cations and anion, a route was developed to replace O atoms with F for tuning the electrical properties of single-crystalline W-$Mg_{0.51}Zn_{0.49}O$ films having a solar-blind 4.43 eV bandgap. The commercially available $ZnF_2$ powder (99.995%, Alfa Aesar) chosen as the source was firstly purified and solidified in order to exclude the possibility of unwanted impurities incorporation and meet the strict requirements by radio-frequency plasma-assisted molecular beam epitaxy (rf-MBE) growth process. Our observations solidly evidence that the incorporation of F does improve the n-type conduction behavior of high-resistance W-MgZnO deep UV components. Accordingly, the UV PD fabricated with $Mg_{0.51}Zn_{0.49}O$:F epitaxial film demonstrated an enhanced photocurrent, photoresponsivity and detectivity, one or two orders of magnitude higher than that of the device fabricated on the undoped film.

## Results

The samples were synthesized on sapphire (0001) substrates by rf-MBE with a base pressure of ~$10^{-10}$ mbar. Reflection high-energy electron diffraction (RHEED) was utilized *in situ* to monitor the whole epitaxial growth process. On oxygen-terminated α-$Al_2O_3$ (0001) surface [shown in Fig. 1(a)], the ultrathin MgO (111) layer provides a good template for subsequent W-MgZnO epitaxy. It should be noted that sharp and streaky RHEED patterns of the highly strained MgO ultrathin layer overlap those of sapphire [Fig. 1(b)], indicating the achievement of an atomically flat surface inherited from the α-$Al_2O_3$ (0001) surface. The elongated, nearly streaky RHEED patterns of the low-Mg-content MgZnO quasi-homo buffer layer [Fig. 1(c)] well accommodates the large mismatch and structural discrepancy between the MgO and the high-Mg-content $Mg_{0.51}Zn_{0.49}O$ epilayer [Fig. 1(d)], which becomes rough due to the much higher Mg content. It can be clearly seen that F doping does not induce any change of $Mg_{0.51}Zn_{0.49}O$ structure except for the rougher surface morphology [Fig. 1(e)].

To confirm the single-crystalline wurtzite structure of the F-doped $Mg_{0.51}Zn_{0.49}O$ layer, X-ray diffraction (XRD) θ–2θ and ϕ–scans were performed. Figure 1(f) shows the XRD θ–2θ curve of the F-doped sample. The peak (41.68°) is attributed to the diffraction from sapphire (006). Diffraction from W-MgZnO:F (002) planes locates at 35.04° obviously shifting to a much larger angle in contrast to that of pure ZnO (34.46°), implying a high Mg content incorporated in the film. Importantly, the appearance of only the (002) related peak without any sign of cubic MgZnO:F confirms the single wurtzite phase, consistently with the *in situ* RHEED findings. The inset in Fig. 1(f) shows an enlarged image of the MgZnO (002) peak, confirming the constant Mg content in the doped and undoped layers. A slight asymmetry in the W-MgZnO:F (002) peak is attributed to the contribution from the low-Mg-content buffer layer underneath. In addition, Fig. 1(g) shows the ϕ–scan of the MgZnO:F (101) plane, which was carried out at $\chi = 60.87°$ [the angle between (002) and (101) planes in a hexagonal system]. Six narrow and sharp peaks with equal 60° intervals can be clearly observed, indicating the common sixfold symmetry of the single wurtzite crystal structure, consistent with the 60° symmetry observed in RHEED patterns.

For comparison, an intrinsic $Mg_{0.51}Zn_{0.49}O$ film, Ga-doped and Al-doped alloy films were also synthesized with the same growth conditions. However, we face the difficulty in obtaining single-crystalline $Mg_xZn_{1-x}O$ alloy films by Ga or Al doping when increasing the Mg content x above a critical level, as other researcher encountered[14,16]. Following the same process as before, the undoped high-Mg-content MgZnO was synthesized, as illustrated in Fig. 2(a,c). After Ga doping, the RHEED patterns show the trend of phase segregation, indicated by the slightly twisted (02) and ($0\bar{2}$) reciprocal spots [Fig. 2(b)]. As Fig. 2(e) shows, the peaks (34.96°, 34.98° and 34.89°) indicate high Mg content incorporated in these films. However, the appearance of additional peaks implies the occurrence of multiple phases after doping. Indeed, it is a challenge to dope such high-Mg-content alloy films via Ga or Al dopants. Introducing cation dopants (Ga or Al) might decrease the Mg solubility in ZnO due to the more competitive bonding between cations (Zn, Mg, Ga or Al) and anions (O). Moreover, the Ga- or Al-doped films all show huge resistance. It is therefore worth evaluating the effect of F doping on tuning the electrical properties and the device performance.

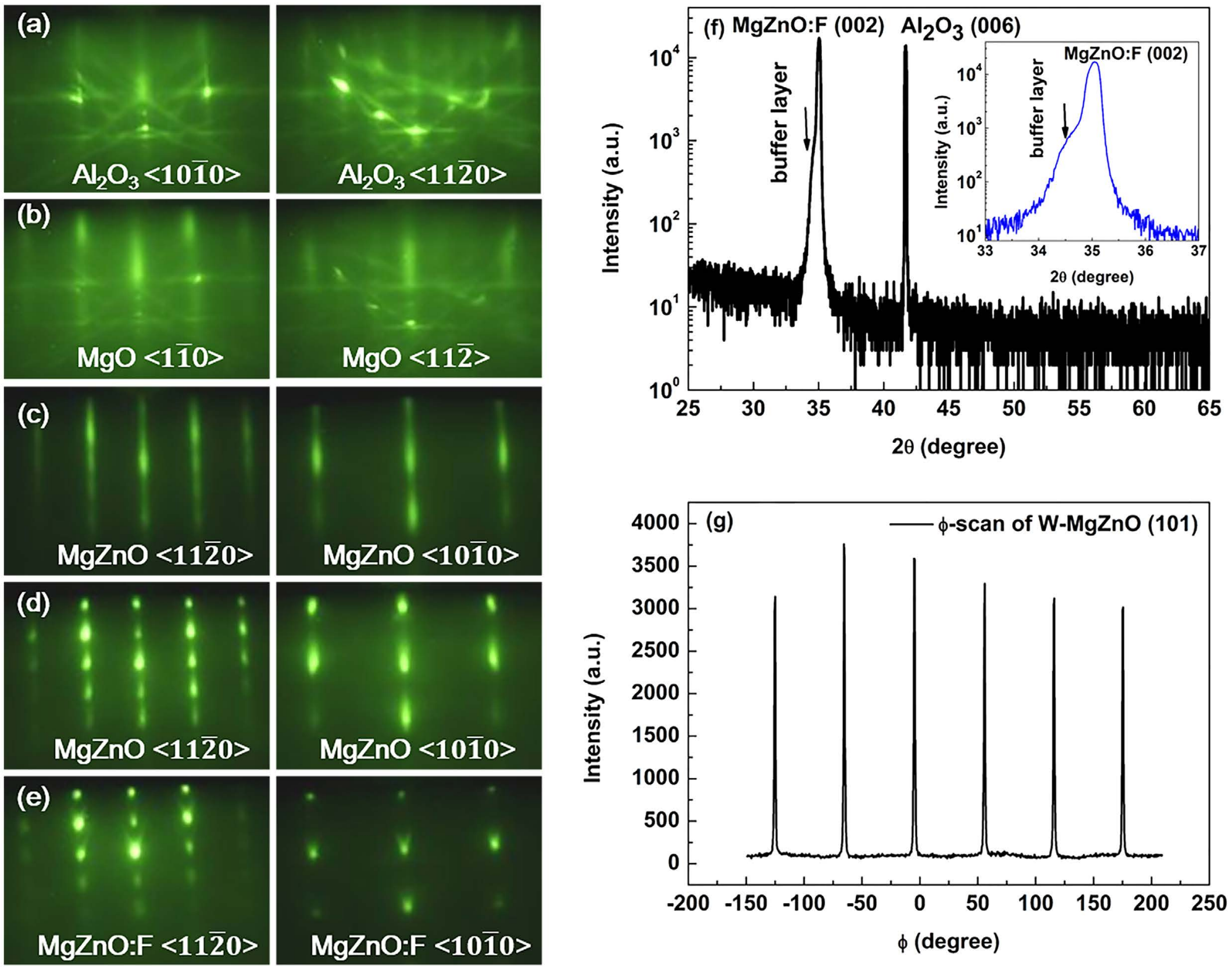

**Figure 1. RHEED patterns with incident electron beams along the $\langle 10\bar{1}0\rangle$ $Al_2O_3$ and $\langle 10\bar{2}0\rangle$ $Al_2O_3$ azimuths, respectively, obtained from $Al_2O_3$ (0001) surface (a); after growth of ultrathin MgO buffer layer at 500 °C (b); after MgZnO buffer growth at 450 °C (c); after MgZnO epitaxial growth at 450 °C (d); and after MgZnO:F epitaxial growth at 450 °C (e).** XRD results of (**f**) θ–2θ scan of W-MgZnO:F (002) on sapphire, and (**g**) ϕ–scan of the W-MgZnO:F (101) plane. The inset shows an enlarged image of the MgZnO (002) peak, confirming the constant Mg content in the doped and undoped layers.

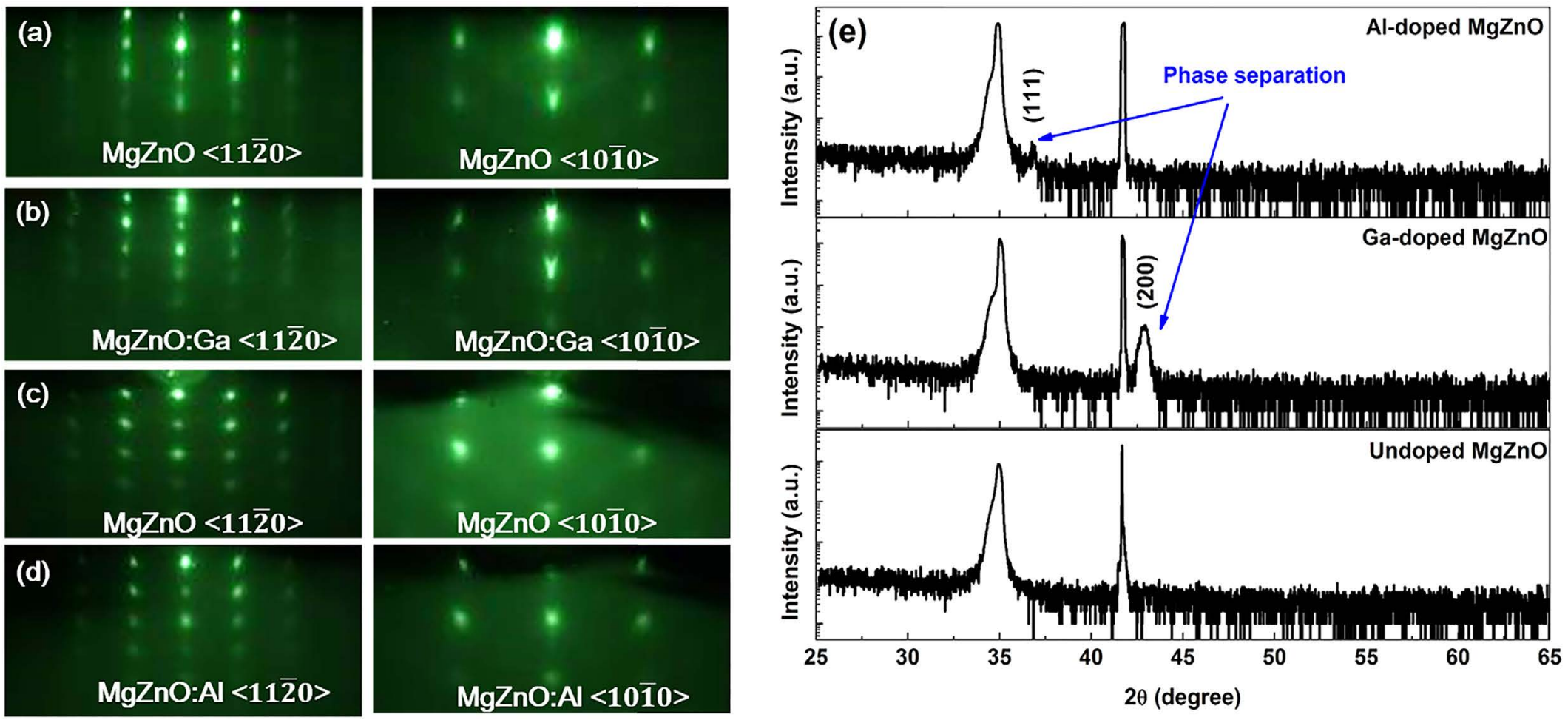

**Figure 2. RHEED patterns obtained from MgZnO epitaxial layer before doping (a,c); after Ga or Al doping (b,d), respectively.** XRD results of θ–2θ scans of undoped MgZnO, MgZnO:Ga and MgZnO:Al layers on sapphire (**e**).

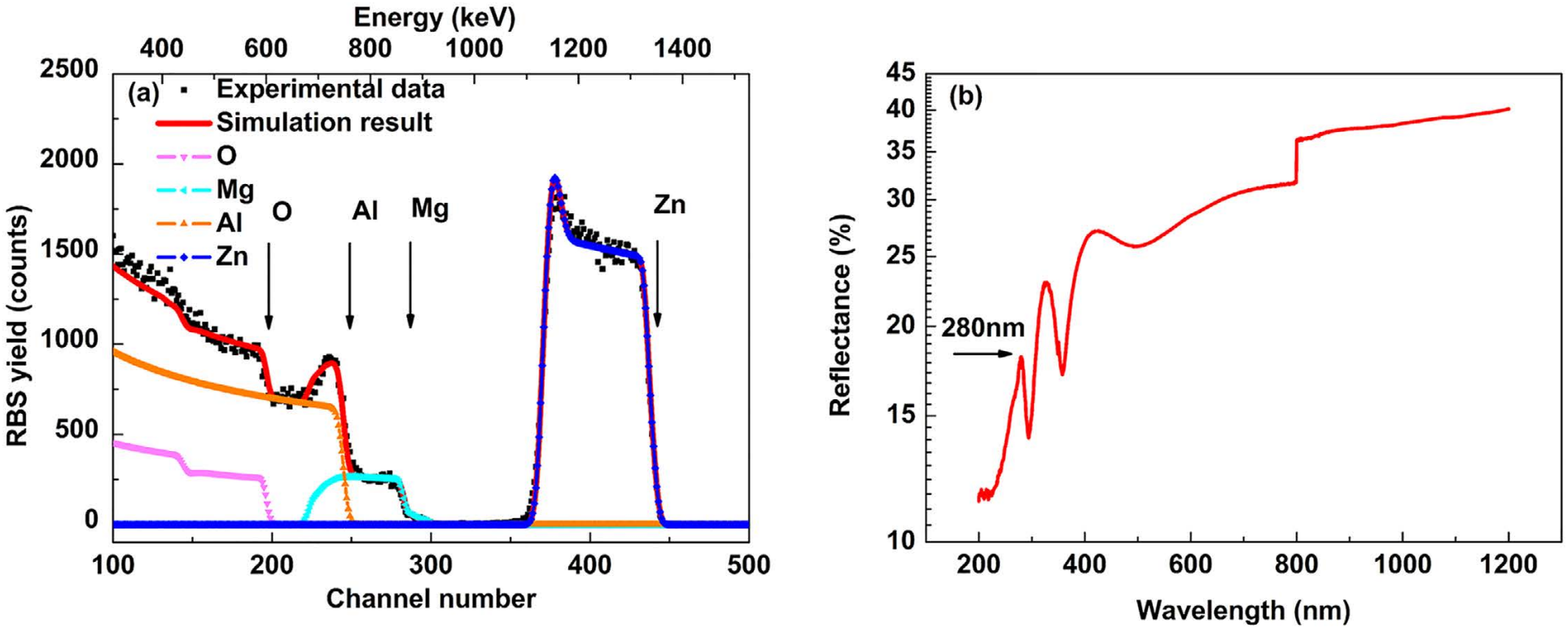

**Figure 3.** (**a**) RBS curves, and (**b**) reflectance spectrum of the same sample measured at room temperature, where the arrow indicates the bandgap position.

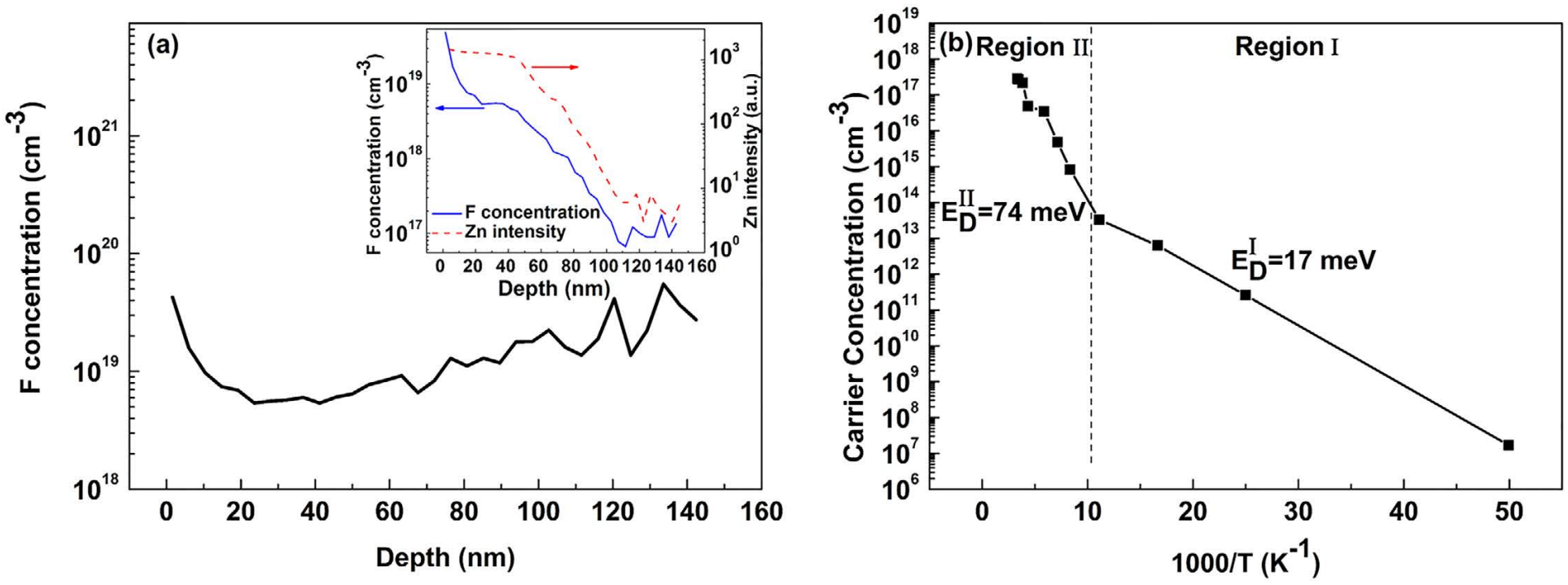

**Figure 4.** (**a**) Normalized F concentration versus depth profile of the $Mg_{0.51}Zn_{0.49}O$:F sample, and (**b**) temperature-dependent Hall results of F-doped $Mg_{0.51}Zn_{0.49}O$ alloy. The inset shows the Zn intensity and the F concentration before normalization.

## Discussion

Figure 3(a) shows a Rutherford backscattering spectrum (RBS) taken from the F-doped sample with 2-MeV $^4He^+$ ions backscattered into the detector at 100° relative to the incident beam direction. Arrows/labels in Fig. 3(a) indicate the channel number at which the backscattering from the corresponding atoms occurs at the surface, except for Al signal, which starts from the sapphire substrate. Note that it is difficult to distinguish F from O in RBS due to their very close mass and high background interference from O. Fitting of the experimental and simulation data [done using SIMNRA code] in Fig. 3(a) reveals the composition of the sample as $Al_2O_3/Mg_{0.36}Zn_{0.64}O/Mg_{0.51}Zn_{0.49}O$:F, consistent with the XRD measurement results.

Room-temperature reflectance spectroscopy was applied to determine the band gap of the sample. As indicated by an arrow in Fig. 3(b), the near-band edge absorption of the $Mg_{0.51}Zn_{0.49}O$:F epilayer is determined to be 280 nm, which corresponds to the optical bandgap of $Mg_{0.51}Zn_{0.49}O$:F (4.43 eV). More importantly, the bandgap of $Mg_{0.51}Zn_{0.49}O$:F is within the solar-blind range, offering the prospects of this material as active layers in the UV-C spectral region (280 nm–200 nm). The abrupt drop at 800 nm in Fig. 3(b) is induced by the exchange action of two different gratings in the testing system.

F concentration versus depth profile in $Mg_{0.51}Zn_{0.49}O$:F using secondary ion mass spectrometry (SIMS) is shown in Fig. 4(a). Due to the charge accumulation into the alloy film, SIMS signals, including these for Zn and F, manifest a monotonically decreasing trend, as illustrated in the inset of Fig. 4(a). Moreover, the data obtained in the range of 110–150 nm decrease a lot, close to the SIMS detection minimum limit, resulting in the huge fluctuation after normalization based on the Zn intensity [Fig. 4(a)]. It should be noted that the undoped MgZnO layer is too isolated to be measured by SIMS under the same conditions. Thus, the F-involved region is about 140 nm thick, in a good accordance with the designed thickness of the F-doped layer (~150 nm). Disregarding to the SIMS uncertainty in the vicinity of the surface as well as fluctuations in the vicinity to the inner interface, the average doping level is estimated to be $1.0 \times 10^{19}$ F/cm$^3$ [Fig. 4(a)].

In order to assess the effect of fluorine incorporation on the tuning of electrical properties, the $Mg_{0.51}Zn_{0.49}O$:F film was characterized by the temperature-dependent Hall measurement (TDH) using the van der Pauw technique in a magnetic field of 10 kG and a temperature range of 20–300 K. Figure 4(b) shows the carrier concentration (n) as the function of the reciprocal temperature, revealing two linear regions, labeled as I and II, respectively. In both regions, the electrical conductivity increases with increasing temperature. In order to determine the donor concentration ($N_D$) and the activation energy ($E_D$), we fitted these data by the least-squares method, assuming the charge neutrality equation for n-type semiconductor containing predominant donors and compensated acceptors:

$$\frac{n\,(N_A - n)}{N_D - N_A - n} = \frac{N_C}{g}\exp\left(-\frac{E_D}{k_B T}\right) \tag{1}$$

where $N_A$, $N_c$, g, $k_B$ and T denote the compensating acceptor concentration, the effective density of states in the conduction band, the donor degeneracy factor (~2), Boltzmann constant and absolute temperature, respectively. The impact from the semi-insulating MgZnO beneath the doped layer was reasonably neglected. The effective electron mass ($m_e$*) in $Mg_{0.51}Zn_{0.49}O$:F was estimated as 0.30 $m_0$ since it is reported the value changes slightly from $m_e{}^*{}_{(ZnO)} = 0.29\,m_0$[18] to $m_e{}^*{}_{(MgO)} = 0.30\,m_0$[19], where $m_0$ is the free electron mass. As a result, the best-fitted values are as follows: for region I, $E_D^I = 17$ meV; and for region II, $N_D = 8.4 \times 10^{18}\,cm^{-3}$, $N_A = 9.7 \times 10^{17}\,cm^{-3}$, $E_D^{II} = 74$ meV. In analogy to the case of GaN:Mg and ZnO:Li/Na[20], the shallow F-donors level $E_D^I$ (17 meV) and the deep one $E_D^{II}$ (74 meV) may be related to the lattice distortions in wurtzite $Mg_{0.51}Zn_{0.49}O$. As we know, ZnO possesses the non-centra symmetric nature of Zn and O sub-lattices. This feature will be more pronounced with more Mg atoms incorporated, reflected by decreased c-lattice parameter confirmed by the XRD results shown in Fig. 1(f) and increased a-lattice parameter[21]. The enhanced polarization along the c-axis could affect the defect level of F in the bandgap. In wurtzite $Mg_{0.51}Zn_{0.49}O$, every substitutional fluorine atom, $F_O$, will averagely form two bonds with two neighbor Mg, and another two bonds with neighbor Zn. There are two different configurations for these $F_O$, i.e. one kind of $F_O$ forming three bonds with two Zn and one Mg below, and one bond with one Mg up, and another kind of $F_O$ forming three bonds with one Zn and two Mg below, but one bond with one Zn up (supplementary information), resulting in two different tetrahedral environments for $F_O$ with different local polarization fields along the c-axis. The F-related energy level may split into two as observed by Hall measurements. The net electron concentration reaches $2.85 \times 10^{17}\,cm^{-3}$ at room temperature, resulting in a resistivity as low as $129\,\Omega \cdot cm$ and decreasing by four orders of magnitude compared to the undoped $Mg_{0.51}Zn_{0.49}O$[11]. It demonstrates that F atoms effectively act as donors by supplying free electrons when they occupy O sites in the hexagonal lattice of $Mg_{0.51}Zn_{0.49}O$.

The impact of the enhanced conductivity on the device performance was evaluated via fabrication of two Schottky type interdigital planar metal-semiconductor-metal (MSM) UV PDs with the undoped and F-doped $Mg_{0.51}Zn_{0.49}O$, respectively. Ti (10 nm)/Au (50 nm) was deposited to form finger electrodes with 5 μm width, 300 μm length, and 5 μm gap, as illustrated in the inset of Fig. 5(a). The well-defined symmetrical rectifying behavior in Fig. 5(b,d) indicates the back-to-back Schottky contacts of the non-alloyed Ti/Au on high-Mg-content films. The dark current of F-doped $Mg_{0.51}Zn_{0.49}O$ UV PD is increased by more than two orders of magnitude compared to that of the undoped one [Fig. 5(b)]. For a Schottky contact in an ideal case, if $E_{00} \approx k_B T$, the thermionic-field emission (TFE) dominates the electronic transport process, which is a combination of thermionic emission (TE) and field emission (FE). $E_{00}$, $k_B$ and T denote the characteristic energy, Boltzmann constant and absolute temperature, respectively. $E_{00}$ is defined as:

$$E_{00} \equiv (q\hbar/2)\,(N/m_e^* \varepsilon_s)^{1/2} \tag{2}$$

Where q, ħ, N, $m_e$* and $\varepsilon_s$ denote the elementary charge, the reduced Planck constant, the carrier concentration, the effective electron mass and the dielectric permittivity, respectively. The carrier concentration N is ~$10^{14}\,cm^{-3}$ and $2.85 \times 10^{17}\,cm^{-3}$ for the undoped and F-doped films, respectively. The dielectric permittivity in $Mg_{0.51}Zn_{0.49}O$ is taken as $9.20\varepsilon_o$ by assuming a linear increase from $\varepsilon_{ZnO} = 8.75\varepsilon_o$[22] to $\varepsilon_{MgO} = 9.64\varepsilon_o$[23] with the Mg-content x in $Mg_xZn_{1-x}O$. As a result, $E_{00}$ is 0.11 meV for the undoped film, which is smaller than the thermal energy $k_B T$ at room temperature (26 meV). However, $E_{00}$ for the doped one increased to a much larger value (6.0 meV) and could be to some extent comparable to the thermal energy. Therefore, the I–V curve is roughly determined by TE model and TFE model for undoped and doped cases, respectively, as illustrated in Fig. 5(c). F doping could narrow the potential barrier and result in larger tunneling currents[24]. Under 254 nm light illumination, the photocurrent of F-doped $Mg_{0.51}Zn_{0.49}O$ UV PD increased by two orders of magnitude compared to that of the undoped one [Fig. 5(d)]. Thus, the photoresponsivity of the device is dramatically enhanced, although sacrificing some contrast ratio.

The time dependence of photocurrent properties was studied by using periodic 254 nm and 365 nm illumination alternatively from a UV lamp [Fig. 6(a,b)]. The 10%–90% rise and decay time are less than 0.12 s for both two PDs, which is the limit of our testing system. The sharp curve [Fig. 6(a)] indicated F doping didn't result in extra defects, especially oxygen vacancies, which could cause persistent photocurrent (PPC)[25]. Figure 6(c) shows the photoresponsivity curve of both devices at a 10 V bias. The cutoff wavelength is 278 nm and 276 nm for undoped and F-doped MgZnO-based devices, respectively, which agrees well with the optical bandgap of these two films. Note that the peak photoreponsivity was enhanced from 0.34 mA/W to 52 mA/W after F doping. (The peak at 320 nm is induced by the exchange action of

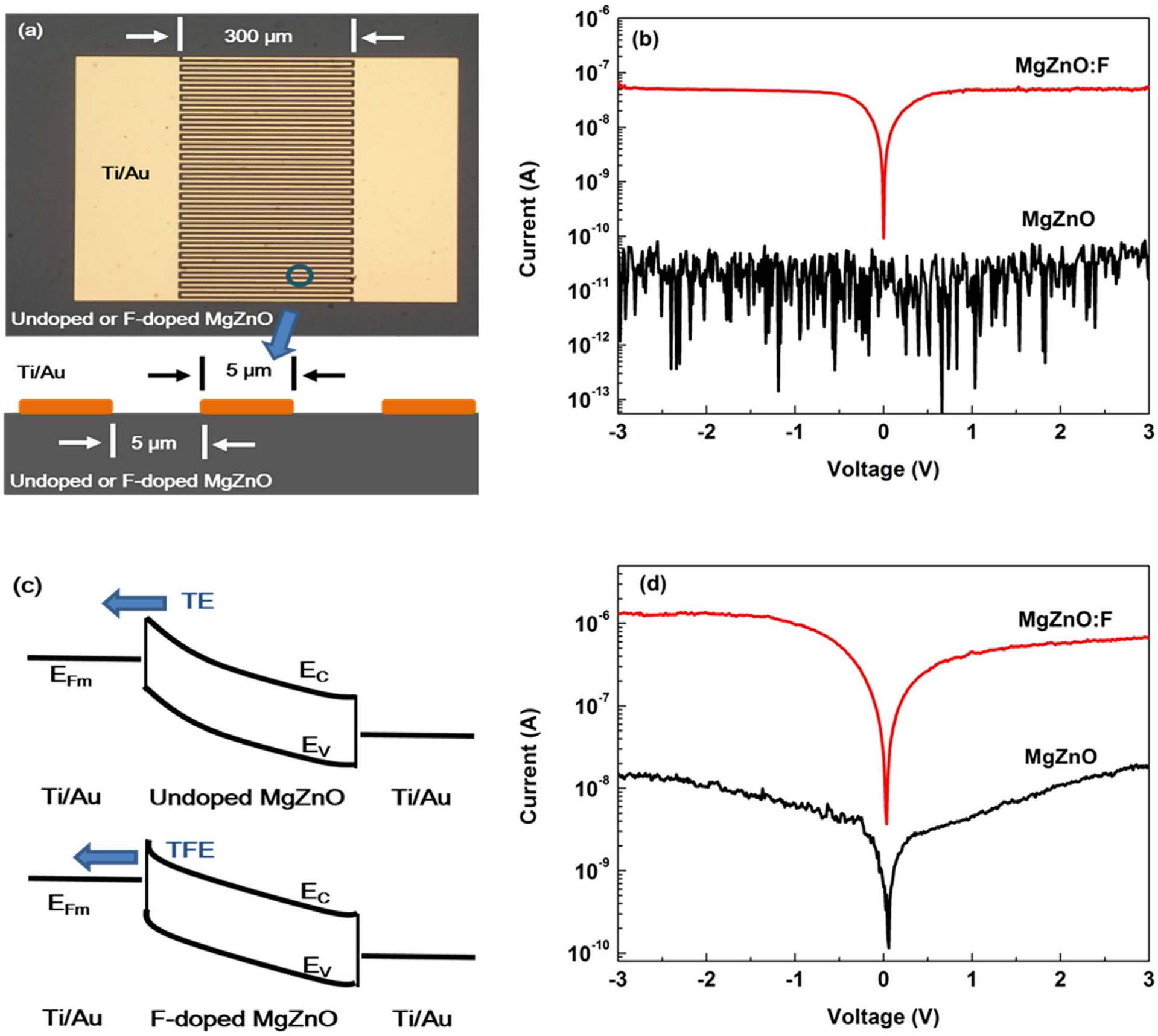

**Figure 5.** (**a**) A schematic diagram of the MSM UV PDs, (**b**) the dark current-voltage (I–V) characteristics of the two identically designed PDs using undoped and F-doped $Mg_{0.51}Zn_{0.49}O$, (**c**) energy diagram under bias, showing the dominant transport mechanism. TE = thermionic emission. TFE = thermionic-field emission, and (**d**) the current-voltage characteristics of the two PDs under 254 nm light illumination.

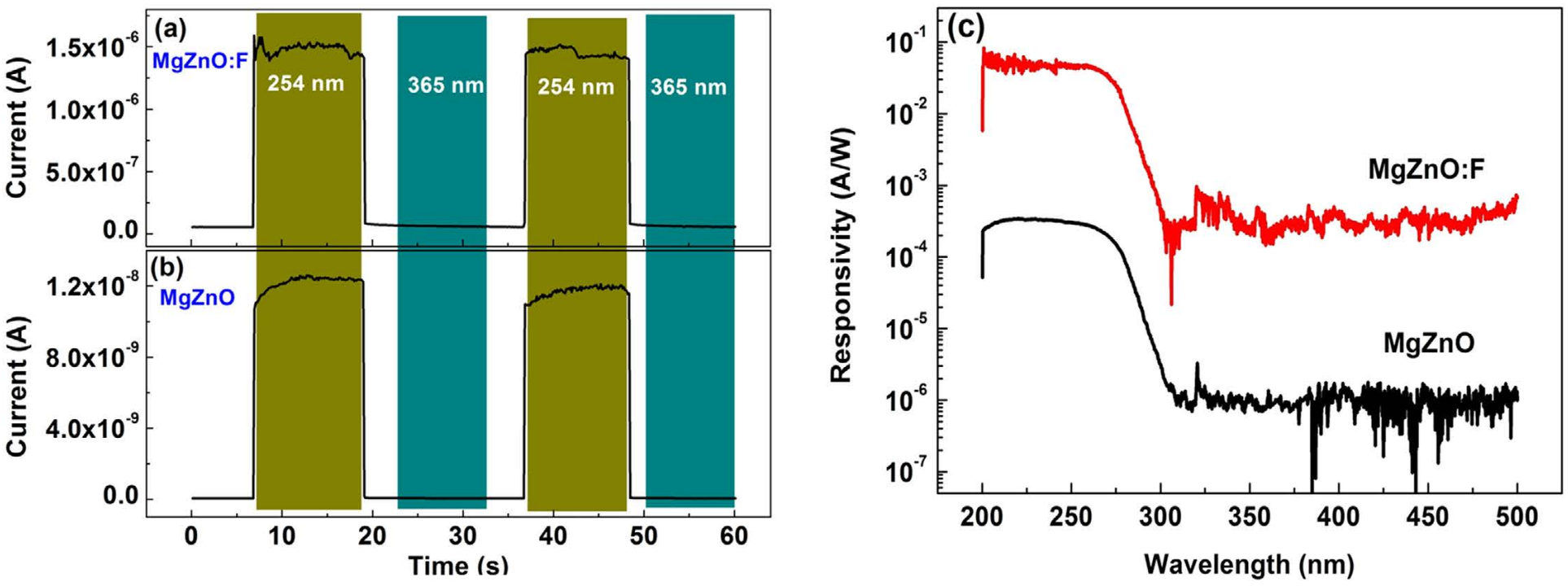

**Figure 6.** **Time-dependent response of the photocurrent of undoped and F-doped MgZnO-based device (a,b) under 3 V bias.** (**c**) Spectral photoresponse of the undoped and F-doped MgZnO solar-blind UV detectors at a 10 V bias.

two gratings in the testing system.) The noise equivalent power (NEP) can be evaluated by NEP = $(4k_BT/R_{dark} + 2qI_{dark})^{1/2}\Delta f^{1/2}/R$, where $R_{dark}$, $I_{dark}$, $\Delta f$ and R denote the differential resistance ($R = dV/dI$) under the bias, the dark current under the bias, the electrical bandwidth and responsivity, respectively. $k_B$,T and q have the same meaning as above. The NEP is determined to be $4.21 \times 10^{-11}$ W/Hz$^{1/2}$ and $2.22 \times 10^{-12}$ W/Hz$^{1/2}$ for the undoped and F-doped device, respectively, under a 10 V bias at room temperature. The detectivity (D*) can be then determined by $D^* = (A\Delta f)^{1/2}/NEP$, where A is the optical active area. The detectiviy D* is $1.89 \times 10^9$ cm Hz$^{1/2}$/W and $3.58 \times 10^{10}$ cm Hz$^{1/2}$/W for undoped and F-doped devices, respectively. The above results indicated F doping would not cause severe quality degeneration of high-Mg-content

MgZnO films, but could enhance the responsivity and detectivity by 1 ~ 2 orders of magnitude. It unambiguously implies that F doping can robustly tune the conductivity of high-Mg-content W-MgZnO films in a controllable way and have a positive impact on the device performance.

To confirm the tuning effect of fluorine, the electrical properties of W-$Mg_xZn_{1-x}O$:F ($0 \leq x \leq 0.3$) thin films were investigated by Hall measurement. The resistivity increased from $5.32 \times 10^{-3}\,\Omega \cdot cm$ to $0.16\,\Omega \cdot cm$ when increasing Mg content x from 0 to 0.3, with fluorine concentration of ~$9.8 \times 10^{19}\,cm^{-3}$ (supplementary information).

In conclusion, the conductive W-$Mg_{0.51}Zn_{0.49}O$:F thin film was prepared with a fluorine concentration of ~$1.0 \times 10^{19}\,cm^{-3}$, lifting the electron concentration from ~$10^{14}\,cm^{-3}$ to $2.85 \times 10^{17}\,cm^{-3}$. The conductivity and photoresponsivity increased by two orders of magnitude. The detectivity was enhanced from $1.89 \times 10^{9}$ cm $Hz^{1/2}$/W to $3.58 \times 10^{10}$ cm $Hz^{1/2}$/W. Two energy levels were revealed after fluorine incorporation, which is tentatively ascribed to built-in electric field discrepancy along the c-axis between two different configurations. The results indicate that F doping can dramatically modulate the electrical properties of high-Mg-content W-MgZnO and improve the UV device performance in solar-blind range, which is of crucial importance to promote the competitiveness of these active components. In addition, the purification and solidification process to $ZnF_2$ powder and their unique advantages, high-purity and solid phase for example, may offer a new approach for fluorine doping attempts in other wide bandgap oxides.

## Methods

F doping into W-$Mg_{0.51}Zn_{0.49}O$ film was realized with a solid anhydrous $ZnF_2$ source by radio-frequency plasma-assisted molecular beam epitaxy (rf-MBE). After degreasing in acetone and ethanol, a sapphire wafer was loaded into the vacuum chamber and thermally cleaned at 750 °C, followed by exposure to active oxygen radicals at 500 °C. An ultrathin unrelaxed cubic MgO buffer layer (~1 nm) was deposited at 500 °C, providing an epitaxial template for the growth of W-MgZnO film. Further, a quasi-homo $Mg_{0.36}Zn_{0.64}O$ buffer layer (~20 nm), a high-Mg-content $Mg_{0.51}Zn_{0.49}O$ epilayer (~80 nm) and a fluorine-doped $Mg_{0.51}Zn_{0.49}O$ epilayer (~150 nm) were subsequently grown at 450 °C, respectively. Following the same growth process, intrinsic $Mg_{0.51}Zn_{0.49}O$ film, Ga-doped and Al-doped alloy films were also synthesized. The K-cell temperature for $ZnF_2$ , Ga and Al was set at 420 °C, 500 °C and 890 °C, respectively. More growth details can be found elsewhere[7,26].

XRD was performed using Cu K$\alpha$ radiation (Empyrean System). RBS system is based on 1 MeV NEC Tandem accelerator. Additional insights into the bandgap were obtained using room-temperature reflectance spectroscopy (Cary 5000 System). SIMS measurements were performed using a Cameca IMS 7 microanalyzer. The electrical properties were characterized by TDH in Lakeshore 7604 system. Semiconductor parameter analyzer (Keithley 6487) was employed for I–V characterization. The spectra response was performed using the SpectraPro-500i (Acton Research Corporation) optical system with a 75 W Xe-arc lamp combined with a 0.5 m monochromator as light source.

## Acknowledgements

This work was supported by the Ministry of Science and Technology of China (Grant nos 2011CB302002, 2011CB302006), the National Science Foundation of China (Grant nos 11174348, 51272280, 11274366, 61204067, 61306011), the Chinese Academy of Sciences, and the Research Council of Norway (Grant no 221668).

## Author Contributions

Z.M. and X.D. conceived and guided the study. L.L. and Z.M. designed the reasearch and guided the work and analysis. L.L. and H.L. conducted the growth and characterization. Y.H. fabricated the devices. A.A., V.V. and A.K. performed the RBS, SIMS and TDH. experiments. L.L. wrote the paper. All authors discussed the results and commented to the manuscript.

## Additional Information

**Supplementary information** accompanies this paper at http://www.nature.com/srep

**Competing financial interests:** The authors declare no competing financial interests.